\documentstyle[11pt,newpasp,twoside,epsf]{article}
\def\emphasize#1{{\sl#1\/}}

\def\edcomment#1{\iffalse\marginpar{\raggedright\sl#1\/}\else\relax\fi}
\reversemarginpar

\begin{document}
\title{An AGN Hertzsprung-Russell Diagram}
 \author{Peter Barthel}
\affil{Kapteyn Astronomical Institute, University of Groningen, The Netherlands}

\begin{abstract}
Detailed examination of the balance between star-formation and nuclear 
activity in AGN and starburst galaxies leads to the composition of an
Hertzsprung-Russell diagram in which possible evolutionary tracks can 
be drawn. It is likely that these tracks also relate to the level of 
obscuration.
\end{abstract}

\section{Introduction}

The question I'd like to address is whether there is a physical reason
behind the resemblance of the cosmic star-formation history diagram, the
`Madau-plot', and the evolving QSO space density diagram (e.g., Shaver
et al.  1996).  Given that the shapes of these plots -- in particular
the $z \sim 2.5$ peaks -- are similar, could it be that star-formation
and nuclear activity in galaxies are more intimately related than we
have believed so far? Rosa Gonzalez Delgado and George Rieke (these
Proceedings) have already presented evidence that such is the case for
Seyfert galaxies -- I will add QSOs and discuss the AGN--starburst
symbiosis within the framework of an Hertzsprung-Russell diagram
analogon.  A more complete account of the issue and a more thorough
presentation of the rationale can be found in Barthel (2001). 

\section{Star-formation and FIR SEDs}

Star-formation is mainly studied in the restframe ultraviolet-blue part
of the spectrum, where clear signatures from young star populations can
be found.  However, such signatures will remain hidden when the
star-formation occurs in a dusty region.  In that case only thermal
reradiation of the the hard ultraviolet photons by the dust can reveal
the obscured star-formation.  Data taken with IRAS and more recently ISO
have beautifully demonstrated the importance of this mechanism: it has
for instance revealed the class of ULIRGs (e.g., Sanders \& Mirabel
1996) and the starburst activity located \emphasize{in between} the
Antennae galaxies NGC\,4038 and NGC\,4039 (Vigroux et al.  1996). 

While normal galaxies display cold dust (heated up to $\sim$20K by the
interstellar radiation field primarily due to the old population),
starburst galaxies display warmer dust ($\sim$50K) related to the
visible or unvisible young star population.  The luminosity of the
latter dust correlates with the (mostly) diffuse radio synchrotron
emission, through the radio-FIR correlation (e.g., Condon 1992).  This
radio-FIR correlation is well known, but in my view not yet appreciated
to its fullest extent -- recall that only radio waves and hard X-rays
can penetrate the inferred walls of extinction! From their far-infrared
SEDs, normal and starburst galaxies are characterized with steeply
rising 25$\mu$m to 60$\mu$m slopes between 1.5 and 3. 

Seyfert galaxies have a third dust component: hot dust peaking at
$\sim$25$\mu$m (De Grijp et al.  1985, Rodr\'{\i}guez Espinosa et al. 
1996).  This radiation is most likely emitted by the circumnuclear torus
dust, being directly exposed to the hard AGN continuum (Rowan-Robinson
1995).  I note in passing that this hot dust component is present at
comparable magnitude in Seyfert galaxies of Type~1 and Type~2 (P\'erez
Garc\'{\i}a et al.  1998), thereby proving the presence of hot,
circumnuclear dust in Type~1 AGN.  This third dust component produces a
flattening of the $\alpha_{25\mu}^{60\mu}$ slope, to values in the range
1.0 to 1.5. 

Not all AGN, however, have $\alpha_{25\mu}^{60\mu}$ indices in that
range.  On one hand there is the Blazar class which radiates
\emphasize{nonthermal} FIR, and can have indices as flat as 0.5.  On the
other hand there are AGN with an unusually luminous star-formation
activity.  The extra-strength 60$\mu$m component related to this
activity steepens the $\alpha_{25\mu}^{60\mu}$ index to values in excess
of 1.5 -- the compact radio-loud quasar 3C\,48 is a prime example
(Canalizo \& Stockton 2000a). 

How general is this enhanced star-formation activity among the QSO
population (radio-loud and radio-quiet)?

\section{Star-forming QSOs and Seyfert galaxies}

Like 3C\,48 mentioned above a few more quasars are known to have strong
ongoing star-formation.  In an infrared color-color diagram (for
instance displaying $\alpha_{25\mu}^{60\mu}$ versus
$\alpha_{60\mu}^{100\mu}$) these objects are found close to the area
occupied by ULIRGs (e.g., Canalizo 2000, Barthel 2001), and so the
question emerges as to whether temporal evolution between the classes
can occur. 

\begin{figure}
\plotfiddle{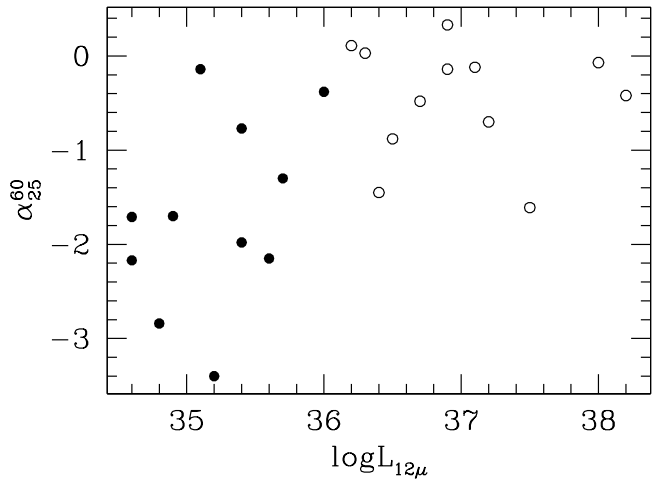}{7cm}{0}{45}{45}{-140}{-120}
\caption{FIR color index $\alpha_{25\mu}^{60\mu}$ as function of
  AGN luminosity log\,L(12$\mu$), for the combined sample of $z<0.4$
  Seyfert galaxies (filled circles) and QSOs (open circles).}
\end{figure}

In our study of the AGN-starburst symbiosis my colleagues and I have
obtained and analysed deep radio (VLA), optical/nearIR (ESO; La Palma,
including Carlsberg Meridian Circle astrometry), as well as improved
IRAS data for a sample of 16 Seyfert galaxies having $z<0.02$, and 27
radio-quiet PG (Palomar-Green) QSOs having $0.02<z<0.4$.  These
complementary samples of active galaxies span a wide, continuous
luminosity range of 3.5 orders of magnitude.  These luminosities are
expressed as L(12\,$\mu$m), and hence predominantly reflect the AGN
strength (e.g., Spinoglio \& Malkan 1989).  The radio-imaging, reaching
noise levels of $\sim 30\,{\mu}$Jy, as well as the optical astrometry
yield AGN positions to a 3$\sigma$ accuracy of $\sim 0.4$\,arcsec.  This
allows comparison of the AGN versus the star-formation driven radio
emission.  The radio data of these -- I stress -- radio-quiet active
galaxies are being combined with far-infrared photometry, yielding
$u$-parameters log\,$S_{60\mu}/S_{6{\rm cm}}$ (see e.g., Condon \&
Broderick 1988), which permit assessment of the relative roles of
nuclear activity and star-formation.  Normal star-forming spirals,
obeying the radio-FIR correlation, have $u$ values in the range
2.4--3.0.  Most Seyfert galaxies have infrared detections (at 25 and
60\,$\mu$m), in contrast to about half of the PG QSOs.  Also the ratios
of the 60\,$\mu$m and blue flux densities were compiled. 

Plotting the infrared color as function of the (AGN) luminosity, we see
from Fig.~1 that the more luminous QSOs (open symbols) are warmer than
the Seyfert galaxies (filled symbols), and we infer that the AGN
luminosity must to some extent be driving the objects' dust
temperatures. 

\begin{figure}
\plotfiddle{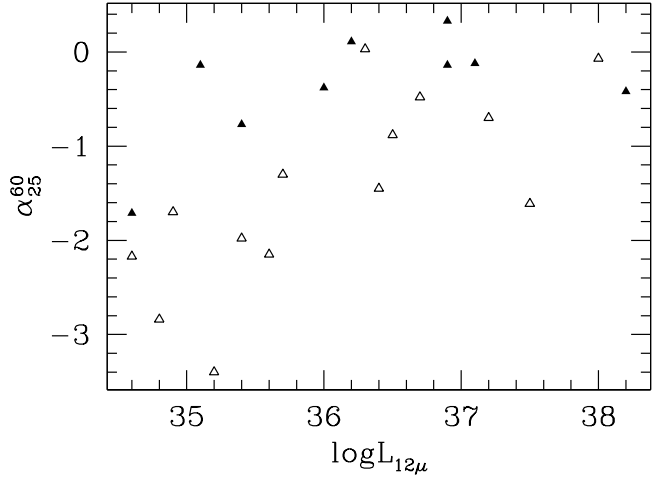}{7cm}{0}{45}{45}{-140}{-120}
\caption{Same objects as in Fig.~1, with relatively strong (open
 symbols) and relatively weak star-formation activity (filled symbols).}
\end{figure}

Analysis of the $u$-parameters in combination with the radio images
subsequently indicates that excess \emphasize{nuclear} radio emission,
and/or fading diffuse radio emission also leads to warmer dust.  This is
the case for both the Seyfert and the QSO class, and implies that an
extra-strength active nucleus in a galaxy with fading starburst activity
also raises the dust temperature.  Fig.~2 shows the same distribution as
Fig.~1, but with filled symbols now representing the
weak-starburst/extra-strength AGN ($u < 2.3$) and the open symbols
representing the objects obeying the radio-FIR correlation ($u \geq
2.3$).  We note a decrease in the $u$-parameter, in the upward direction
in Fig.~2: for all (AGN) luminosities the dust becomes warmer with
fading star-formation.  The 60\,$\mu$m over B-band flux density ratios 
are in good agreement: a strong decrease is seen, in the
upward direction of Fig.~2, implying fading of the cool star-formation
related dust component for a given (AGN) B-band luminosity.

\section{An Hertzsprung-Russell diagram equivalent?}

The objects in Figures 1 and 2 make up a -- be it wide -- diagonal strip
running from bottom left to top right.  This strip in turn can be
separated in a lower part where star-formation is relatively important,
and an upper part where star-formation is relatively unimportant. 
Classical ULIRGS such as Arp\,220 and Mkn\,231 are characterized with
rather cool dust (see the beginning of Section~3) and fall under this
lower strip.  When rotated by $90^{\circ}$, the figures display
luminosity versus temperature, and can thus be considered as the active
galaxy equivalent of the classical Hertzsprung-Russell diagram.  If the
question as for an evolutionary connection between the strongly and the
weakly star-forming AGN is a valid one, then also the connection between
the AGN and the ULIRGs should be taken seriously (cf.  Sanders et al. 
1988).  Age dating (e.g., Canalizo \& Stockton 2000b, Canalizo 2000) may
permit to draw the evolutionary tracks.  JCMT observations of the
molecular gas in QSO hosts are currently being analyzed by our group to
test the evolutionary scenario.  I conclude that a substantial fraction
of AGN displays strong star-formation activity, and note that such is
probably not restricted to radio-quiet AGN (e.g., Aretxaga et al.  2001).

\section{Implications}

If an appreciable, more or less constant fraction of strongly
star-forming galaxies develops and reveals an AGN after and/or during
the process of intense circumnuclear star-formation, the similarity of
the QSO space density and the star-formation history diagram may not be
coincidental.  Given that the latest versions of the latter (e.g.,
Calzetti 2001) suggest a non-declining star-formation rate for redshifts
$z=2$ to 5, and given the submm indications for even more distant
starburst activity (e.g., Dunlop 2001), the hunt for extreme redshift
QSOs remains at order.

\section{Conclusions}

The 60\,$\mu$m luminosity, when normalized with the 25\,$\mu$m (or the
12\,$\mu$m), the blue optical or the radio luminosity, permits
assessment of the absolute and relative strength of star-formation and
nuclear activity in active galaxies and quasars.  The FIR temperature
can be combined with measures of the bolometric luminosity to yield an
intriguing AGN Hertzsprung-Russell diagram, which among other things
suggests that star-formation plays an important role in many AGN.  Such
photometric ratios can be obtained in a straightforwardly manner for the
faint distant objects to be measured in large quantities with upcoming
space-infrared missions, such as SIRTF, ASTRO-F and Herschel.

\acknowledgements

Travel support from the Leiden Kerkhoven-Bosscha Fonds and the EU
Network "Probing the Origin of the Extragalatic Background" is
gratefully acknowledged.  I furthermore acknowledge a long collaborative
effort with Bob Argyle, Jeroen Gerritsen, Magiel Janson, Johan Knapen,
David Sanders and Dick Sramek.

\end{document}